\documentclass[sigconf]{acmart}

\usepackage{listings}
\usepackage[capitalise]{cleveref}
\usepackage{amsfonts}
\usepackage{siunitx}

\definecolorset{RGB}{tol}{}{blue,68,119,170;cyan,102,204,238;green,34,136,51;yellow,204,187,68;red,238,102,119;purple,170,51,119;grey,187,187,187}

\newif\ifbuild
\buildfalse
\ifbuild
\usepackage{tikz}
\usepackage{pgfplots}
\pgfplotscreateplotcyclelist{bright}{%
	tolblue,every mark/.append style={fill=tolblue!80!black},mark=*\\%
	tolcyan,every mark/.append style={fill=tolcyan!80!black},mark=square*\\%
	tolgreen,every mark/.append style={fill=tolgreen!80!black},mark=triangle*\\%
	tolyellow,every mark/.append style={fill=tolyellow!80!black},mark=star\\%
	tolred,every mark/.append style={fill=tolred!80!black},mark=diamond*\\%
	tolpurple,every mark/.append style={solid,fill=tolpurple!80!black},mark=pentagon*\\%
	tolgrey,every mark/.append style={solid,fill=tolgrey!80!black},mark=10-pointed star\\%
}

\usetikzlibrary{external}
\pgfplotsset{compat=1.18, `
	/pgfplots/bar  cycle  list/.style={/pgfplots/cycle  list={%
			{tolblue!20!black,fill=tolblue,mark=none},%
			{tolcyan!20!black,fill=tolcyan,mark=none},%
			{tolgreen!20!black,fill=tolgreen,mark=none},%
			{tolyellow!20!black,fill=tolyellow,mark=none},%
			{tolred!20!black,fill=tolred,mark=none},%
			{tolpurple!20!black,fill=tolpurple,mark=none},%
			{tolgrey!20!black,fill=tolgrey,mark=none},%
		}
	},
	cycle list name=bright,
	table/search path={../data},
}
\fi

\newcommand{\nextsim}{neXtSIM-DG}

\newcommand{\vb}{\mathbf{v}}

\newcommand{\sigmab}{\boldsymbol{\sigma}}

\renewcommand{\div}{\operatorname{div}}

\newcommand{\DB}[1]{{\color{tolblue}#1}}
\newcommand{\DR}[1]{{\color{tolred}#1}}
\newcommand{\DG}[1]{{\color{tolgreen}#1}}

\begin{document}

\title{Towards a GPU-Parallelization of the \nextsim\ Dynamical Core}

\author{Robert Jendersie}
\email{robert.jendersie@ovgu.de}
\affiliation{%
	\institution{Institute for Simulation and Graphics and Institute of Analysis und Numerics, Otto-von-Guericke University}
	\streetaddress{Universitätsplatz 2}
	\city{Magdeburg}
	\country{Germany}
	\postcode{39106}
}

\author{Christian Lessig}
\email{christian.lessig@ecmwf.int}
\affiliation{%
	\institution{European Centre for Medium-Range Weather Forecasts}
	\streetaddress{Robert-Schuman-Platz 3}
	\city{Bonn}
	\country{Germany}
	\postcode{53175}
}

\author{Thomas Richter}
\email{thomas.richter@ovgu.de}
\affiliation{%
	\institution{Institute of Analysis und Numerics, Otto-von-Guericke University}
	\streetaddress{Universitätsplatz 2}
	\city{Magdeburg}
	\country{Germany}
	\postcode{39106}
}

\begin{abstract}
	The cryosphere plays a significant role in Earth's climate system. Therefore, an accurate simulation of sea ice is of great importance to improve climate projections. To enable higher resolution simulations, graphics processing units (GPUs) have become increasingly attractive as they offer higher floating point peak performance and better energy efficiency compared to CPUs. However, making use of this theoretical peak performance, which is based on massive data parallelism, usually requires more care and effort in the implementation. In recent years, a number of frameworks have become available that promise to simplify general purpose GPU programming. In this work, we compare multiple such frameworks, including CUDA, SYCL, Kokkos and PyTorch, for the parallelization of \nextsim, a finite-element based dynamical core for sea ice. We evaluate the different approaches according to their usability and performance.
\end{abstract}



\acmConference[PASC]{Platform for Advanced Scientific Computing}{June 03--05,
	2024}{Zurich, Switzerland}

\maketitle

\section{Introduction}
Simulations are critical to understand the effects of climate change and to enable stakeholders to mitigate its impact on societies and individuals~\cite{Jakob2023}.
An important part of the Earth's climate system is the cryosphere, which impacts in particular long term processes.
The Scale-Aware Sea Ice Project (SASIP)\footnote{\url{https://sasip-climate.github.io}} is currently developing the novel sea ice model \nextsim\ for climate simulations that aims to both improve the representation of the physical processes and the efficiency and accuracy of the numerical implementation.
  
An important factor for the fidelity and reliability of climate simulations is the horizontal resolution, with \qty{1}{km}, convection-resolving simulations being targeted for the next generation of models~\cite{Stevens2019,Bauer2021}.
The resulting computations necessitate exascale HPC systems with significant GPU-based accelerators~\cite{Schaer2020, Bauer2021a}.
Substantial efforts have hence been devoted to porting components of existing climate models to the GPU~\cite{Ikuyajolu2023, Sauer2020, Cao2023, Sun2023}.

In line with these developments, SASIP aims to substantially increase the resolution of the sea-ice component of coupled climate models.
Its dynamical core builds on a viscous plastic sea ice model with a discretization based on a higher order discontinuous/continuous Galerkin method~\cite{Richter2023}.
Computationally, this leads to a large number of identical per-element operations. 
This data parallelism makes the computations well suited for GPUs, which are built around a data parallel processing model.

While performance is paramount, there are other factors of concern for the implementation. Being viable also for future hardware developments and having software support as well as the ease-of-use will play an important role in its adoption and long-term use. With these aspects in mind, we survey the current landscape of general purpose GPU programming frameworks for the parallelization of finite-element/finite-volume codes such as \nextsim.

CUDA remains very popular and is the de facto standard for general purpose GPU programming but only works on NVIDIA hardware.
Furthermore, it typically implies a dedicated GPU implementation whose development and performance tuning requires substantial effort.
For these reasons, different options have emerged that promise greater flexibility and usability.
Directive-based GPU programming frameworks such as OpenMP and OpenACC typically require minimal changes to the code but also provide less control and options for performance optimizations when compared to CUDA.
SYCL~\cite{syclmain} and Kokkos~\cite{Trott2022} are frameworks designed for heterogeneous computing that allow to target different compute hardware.
A major challenge for these is to attain close-to-optimal performance across different hardware platforms and in particular to efficiently exploit GPU specific features such as shared memory or tensor cores. 
Another recent alternative are libraries such as jax~\cite{jax2018github} and PyTorch~\cite{Paszke_PyTorch_An_Imperative_2019} that were primarily developed for machine learning. Their backends consist of high-performance linear algebra libraries that target different hardware and that include a compiler to map computations efficiently onto these.
Examples for these backends include XLA~\cite{xlarepo}, Triton~\cite{Tillet2019} and TensorRT~\cite{tensorrtmain}.

To compare these options, we use them to port an important part of the \nextsim\ dynamical core to run on GPU. 
Our results show that SYCL is still immature, suffering from an unreliable toolchain. Dedicated CUDA remains the best option for speed, while Kokkos provides comparable performance and greater flexibility. PyTorch is currently not a viable alternative to hand written C++ code, but the new compiler TorchInductor shows promise.


The remainder of the paper is structured as follows. 
In \cref{sec:model} we give a brief overview of the \nextsim\ dynamical core. 
The different GPU implementations of it are detailed in the subsequent \cref{sec:implementation}.
Results are presented in \cref{sec:experiments} and directions for future work and a summary is provided in \cref{sec:conclusion}.

\section{Model Description}
\label{sec:model}

The viscous plastic sea ice model by Hibler~\cite{Hibler1979} consists of advection equations for the ice height $H$ and ice concentration $A$
\begin{equation}\label{si:ad}
  \partial_t H + \operatorname{div}\,(\vb H) = S_H,\quad
  \partial_t A + \operatorname{div}\,(\vb A) = S_A,
\end{equation}
and the momentum equation for the two-dimensional velocity field 
\begin{equation}\label{si:momentum}
  \rho_\text{ice}H\partial_t \vb = \div\,\sigmab(\vb,A,H) + F,
\end{equation}
with the ice density $\rho_\text{ice}$, the stress tensor $\sigma$ and $F$ combining all external forcings. Various material models are discussed in the literature~\cite{Feltham2008} but we focus on the classical viscous-plastic model. Since implicit solvers for this highly non-linear model are costly~\cite{MehlmannRichter2017}, one often uses an explicit iteration in the sense of a pseudo-time stepping iteration, namely the mEVP solver~\cite{Bouillon2013}
\begin{equation}\label{si:mevp}
  \begin{aligned}
    \sigmab^{(p)} &= \frac{\alpha}{1+\alpha}\sigmab^{(p-1)} + \frac{1}{1+\alpha}
    \sigmab_{vp}(\vb^{(p-1)},A,H),\\
    \sigmab_{vp}(\vb,A,H) &= \eta(\nabla \vb+\nabla \vb^T)
    +\zeta\operatorname{div}(\vb)I- \frac{P}{2}I,
  \end{aligned}
\end{equation}
where the viscosities $\eta,\zeta$ and depend on velocity $v$, ice height $H$ and ice concentration $A$, while the ice strength $P$ depends only on $H$ and $A$. The parameter $\alpha>0$ controls the stability and the speed of convergence.
The typical solution procedure is thereby as follows: The advection equation~\eqref{si:ad} is solved with a large time step. The momentum equation~\eqref{si:momentum} and mEVP iteration~\eqref{si:mevp} are then subcycled with a smaller step size. Usually, more than 100 substeps are performed per advection step. The main effort in each case is the evaluation of the non-linear material law, given by $\sigma(\vb,A,H)$. Details on the discretization and solution method are given in~\cite{Richter2023}.

\subsection{Discretization}

The \nextsim\ dynamical core~\cite{Richter2023} discretizes equations~\eqref{si:ad}-\eqref{si:mevp} on quadrilateral meshes in spherical coordinates. Discretization of the advection equations is by higher order discontinuous Galerkin methods using a standard upwind scheme. In time, higher order explicit Runge-Kutta methods are used. The velocity is discretized using higher order continuous finite elements and the stress variable is represented in the gradient of the velocity space, which results in a tensor-valued discontinuous Galerkin method.

The mesh is topologically fully structured and the mesh elements are mapped from a reference element onto the computational mesh for better alignment with coastlines and a more equal mesh spacing. This isoparametric element map comes from the same finite element space as the velocity. Details are again given in~\cite{Richter2023}.

\section{Implementation}
\label{sec:implementation}

Our work uses the C++ CPU implementation of the \nextsim dynamical core~\cite{Richter2023} as basis for the GPU port and as reference for performance evaluation.
The CPU implementation leverages the linear algebra library Eigen~\cite{eigenweb}, which is highly optimized and, e.g., exploits CPU vector units. 
Computations consist mostly of a large number of matrix-vector products and component-wise vector operations with small vectors, e.g. of size $8$, corresponding to per-element operations in the finite element discretization.
Since the vector sizes are known at compile time, the implementation benefits greatly from Eigen's template-based design. 
In particular, fixed sized matrices do not require dynamic allocations and operations involving such matrices can be fully unrolled. Furthermore, the use of expression templates in Eigen eliminates unnecessary temporary variables in expressions involving multiple operations.

\begin{table}
	\begin{tabular}{cccc}
		\hline
		& serial [\unit{s}] & OpenMP [\unit{s}] & speedup \\
		\hline
		advection & 188.32 & 34.34 & 5.48 \\
		\hline
		strain & 918.84 & 185.28 & 4.96 \\
		stress & \textbf{1741.43} & \textbf{206.36} & 8.44 \\
		divergence & 1023.07 & 170.81 & 5.99 \\
		velocity & 728.80 & 85.73 & \textbf{8.50} \\
		\hline
		total & 4653.2 & 699.75 & 6.65 \\
		\hline
	\end{tabular}
	\caption{Runtime of 120 time-steps of the simulation with \num{2.6d5} elements on a 10 core CPU. Except for the advection, all major computations are part of the mEVP iteration which performs 100 sub-steps in each time-step.}
	\label{tab:runtimes-detailed}
\end{table}

In \cref{tab:runtimes-detailed} we show computation times for the different parts of the dynamical core. 
As can be seen there, the mEVP iteration (middle lines of the table from "strain" to "velocity") takes most of the time, with the stress update in particular being the single most expensive part. 
Since the stress update also scales well with more cores, it is well suited as computational unit for the evaluation of the different GPU programming frameworks.
The computations for the stress update is provided as pseudo code in \cref{code:stress} and the original C++ can be found in \cref{sec:appendixcode}. While the code is in principle generic in the floating-point type, we use \lstinline|double| in our experiments unless otherwise stated.

\lstset{language=C++, 
	tabsize=4, 
	basicstyle=\small, 
	stringstyle=\ttfamily,
	escapeinside={@}{@}}
\lstMakeShortInline[columns=fixed]|
\begin{figure*}
  \begin{lstlisting}[mathescape=true,numbers=left, caption={
        Implementation of the mEVP iteration equation~\eqref{si:mevp}.
      Stress and strain tensor components $\DB{S^{11},S^{12},S^{22},E^{11},E^{12},E^{22}}\in\mathbb{R}^{N\times n_S}$ are stored as matrices where $N$ is the number of elements and $n_S$ the number of local DOF's in the stress space. Ice height and concentration are denoted as $\DB{H,A}\in\mathbb{R}^{N\times n_A}$, where $n_A$ is the number of local DOFs in the advection space. By $\DB{H_{i,*}}\in\mathbb{R}^{n_A}$ (and similar for the stress and the strain) we denote the local row vector of the DOFs belonging to element $i$. The matrices $\DR{\text{PSI}\langle n_A\rangle}\in\mathbb{R}^{n_A\times n_G}$ are given at compile time and they evaluate the dG functions in the Gauss points with $n_G$ being the number of Gauss points. The scalars $P^\star, \Delta_\text{min} \in \mathbb{R}$ are physical parameters and constant for the duration of the simulation. The matrices $\DG{M_i^{-1}}\in\mathbb{R}^{n_S\times n_G}$ a pre-assembled and stored for each element. They represent the local inverse mass matrix scaled with the weights coming from the transformation of the mesh elements. By ``$*$'' we denote the element-wise Hadamard product of matrices.}, captionpos=b, label=code:stress]
void StressUpdateHighOrder(Matrix<N,$n_S$>& $\DB{S^{11}}$, Matrix<N,$n_S$>& $\DB{S^{12}}$, Matrix<N,$n_S$>& $\DB{S^{22}}$,
                           const Matrix<N,$n_S$>& $\DB{E^{11}}$, const Matrix<N,$n_S$>& $\DB{E^{12}}$, const Matrix<N,$n_S$>& $\DB{E^{22}}$,
                           const Matrix<N,$n_A$>& $\DB{H}$, const Matrix<N,$n_A$>& $\DB{A}$, double $\DB{\alpha}$) {
  for ($i=0$; $i < N$; $++i$) @\label{line:mainloop}@ { # in parallel 
      Vector<$n_G$> $h$     = $\max\{0,\DB{H_{i,*}}\DR{\text{PSI}\langle n_A\rangle}\}$
      Vector<$n_G$> $a$     = $\min\{1,\max\{0,\DB{A_{i,*}}\DR{\text{PSI}\langle n_A\rangle}\}\}$
      Vector<$n_G$> $e^{11}$ = $\DB{E^{11}_{i,*}}\DR{\text{PSI}\langle n_S\rangle}$
      Vector<$n_G$> $e^{12}$ = $\DB{E^{12}_{i,*}}\DR{\text{PSI}\langle n_S\rangle}$
      Vector<$n_G$> $e^{22}$ = $\DB{E^{22}_{i,*}}\DR{\text{PSI}\langle n_S\rangle}$
      
      Vector<$n_G$> $P$     = $P^\star\cdot h * \exp\big(-20(1-a)\big)$
      Vector<$n_G$> $D$     = $\big(\Delta_\text{min}^2 + \frac{5}{4}(\DB{E^{11}_{i,*}}*\DB{E^{11}_{i,*}}+\DB{E^{22}_{i,*}}*\DB{E^{22}_{i,*}}) + \frac{3}{2}\DB{E^{11}_{i,*}}*\DB{E^{22}_{i,*}}+\DB{E^{12}_{i,*}}*\DB{E^{12}_{i,*}}\big)^\frac{1}{2}$
      Vector<$n_G$> $P_D$ = $P/D$
      
      $\DB{S^{11}_{i,*}}$ = $(1-\DB{\alpha}^{-1})\DB{S^{11}_{i,*}} + \DB{\alpha}^{-1}\DG{M^{-1}_i}\big( P_D * (\frac{5}{8} e^{11} + \frac{3}{8} e^{22}) - \frac{1}{2} P\big)$
      $\DB{S^{12}_{i,*}}$ = $(1-\DB{\alpha}^{-1})\DB{S^{12}_{i,*}} + \DB{\alpha}^{-1}\DG{M^{-1}_i}\big( P_D * \frac{1}{4} e^{12}\big)$  @\label{line:mapprod}@
      $\DB{S^{22}_{i,*}}$ = $(1-\DB{\alpha}^{-1})\DB{S^{22}_{i,*}} + \DB{\alpha}^{-1}\DG{M^{-1}_i}\big( P_D * (\frac{5}{8} e^{22} + \frac{3}{8} e^{11}) - \frac{1}{2} P\big)$
  }
}      
\end{lstlisting}
\end{figure*}

\subsection{OpenACC and OpenMP}
A simple approach for moving computations to the GPU is to use a directive based model like OpenACC or OpenMP. Then only small or no changes to the code are required.
Targeting C, C++ and Fortran, both OpenACC or OpenMP define directives to annotate loops. These instruct the compiler to offload the computations onto the GPU. 

OpenACC or OpenMP differ in how the parallel execution is described. OpenMP is \emph{prescriptive}, meaning that the programmer has to detail how a loop should be parallelized. On the other hand, OpenACC provides a simpler \emph{descriptive} directive that leaves more decisions to the compiler. See \cite{Usha2020} for more details on the differences between both approaches. 
In practice, OpenACC tends to give better performance~\cite{Usha2020, Dukic2023}.
However, it has more limited compiler support and except for basic support in GCC, OpenACC can only be used with experimental and commercial compilers that mostly target NVIDIA hardware. 

To accelerate our code, we tried three different compilers: GCC-12.2 and NVIDIA HPC-23.5 with support for both OpenMP and OpenACC, and Clang-16.0 which currently only supports OpenMP. 
However, we quickly found all three compilers to be ill-suited for our purposes. 
The NVIDIA compiler refused to compile Eigen code, while GCC and Clang either crashed during compilation or produced a broken program that would crash once executed. 
A further issue with all but the newest Clang was, that the compilers yielded incorrect memory transfer without a warning or error message. 
In particular, objects which are not trivially copy-able, such as Eigen matrices with at least one dynamic dimension, are not captured properly.
While directives are provided to manually specify the needed buffers, this is cumbersome to do for the complicated template-based Eigen types in use. 
Furthermore, this voids the main advantage of the directive based approach, namely its simplicity.
Use of OpenMP and OpenACC was therefore not pursued further.

\subsection{CUDA}
The de facto standard for general purpose GPU programming remains CUDA~\cite{cudaguide}, a C++ based language and API developed by NVIDIA. CUDA has a mature ecosystem and gives low level access to the GPU, enabling highly optimized code. However, CUDA is limited to NVIDIA hardware.

Starting with version 3.3, Eigen has limited support for CUDA. In particular, fixed sized matrices can be used in CUDA kernels.
This allows us to use the code from \cref{code:stress} largely unchanged. While Eigen's manually vectorized code paths need to be disabled to make it work, we still benefit from Eigen's other features mentioned above such as compile time size-based optimizations and expression templates. To use Eigen with CUDA, we have to ensure that the required data is available on GPU. For dynamic buffers like $\DB{S^{11}}$ in the code listing, we simply allocate memory manually and copy it as needed before and after the kernel invocation. Inside the CUDA kernel, an \lstinline|Eigen::Map| is constructed with
\begin{lstlisting}[mathescape=true]
auto $\DB{B}$ = Map<Matrix<$N$,$n$>>(buf, $N$, $n$);
\end{lstlisting}
which provides the same interface as the original matrix. For compile time matrices such as $\DR{\text{PSI}}$ we use constant memory instead. Advantages of constant memory are that no manual memory management is required, memory access can be faster since it goes through a special cache, and the actual coefficients are available to the compiler, enabling further optimizations. In the original C++ CPU code, the constant matrices are defined as static class members with explicit template specialization to enable selection of the proper matrix for the specified dG-degree at compile time. Since static member variables are not supported in CUDA, we instead declare separate variables and utilize \lstinline|if constexpr| to achieve the same flexibility for generic code:
\begin{lstlisting}[mathescape=true]
__constant__ constexpr T PSI_1_1[1] = {1.0};
template<int $n$, int $n_G$> __device__ auto $\DR{\text{PSI}}$(){
	if constexpr ($n$ == 1 && $n_G$ == 1) {
		return Map<const Matrix<T,1,1>>(PSI_1_1);
	}
}
\end{lstlisting}
Another change to consider when using Eigen on GPU is to set the index type to \lstinline|int|, since the default 64-bit integers are only emulated on GPUs.

\begin{table}
  \begin{tabular}{ccc}
		\hline
		optimization & time [\unit{s}] & speedup \\
		\hline
		CUDA baseline & 0.366 & 1.0 \\
		CUDA shared memory & 0.371 & 0.99 \\
		CUDA column-major & 0.419 & 0.87 \\
		CUDA on-the-fly map & 0.323 & 1.13 \\
		\hline
		AdaptiveCPP baseline & 0.466 & 1.0 \\
		AdaptiveCPP shared memory & 0.531 & 0.88 \\
		AdaptiveCPP on-the-fly map & 0.375 & 1.24 \\
		\hline
		Kokkos baseline & 0.522 & 1.0 \\
		Kokkos shared memory & 0.551 & 0.95 \\
		Kokkos on-the-fly map & 0.386 & 1.35 \\
		\hline
	\end{tabular}
	\caption{Total time spend on the stress computation for the different implementations on an A100. Each modification is tested independently and speedup is relative to the respective baseline.}
	\label{tab:optimization}
\end{table}
We tried a number of optimizations to speed up the Eigen CUDA code, the results of which are shown in \cref{tab:optimization}. Frequently, the bottleneck on the GPU is memory access. One remedy is the manual use of the L1-cache, called \emph{shared memory} in CUDA. Shared between all threads in a thread block, it can significantly speed up reads of data that is needed multiple times and by multiple threads or when scattered memory reads/writes are necessary. In \cref{code:stress}, the only data that are used multiple times and by multiple threads are the $\DR{\text{PSI}}$ matrices.
Only minor changes to the code are needed to load the $\DR{\text{PSI}}$ matrices into shared memory before use. However, we see no benefit from this change as shown in \cref{tab:optimization}. Access through the constant cache is just as fast for the compile time matrices. Shared memory would therefore only be worth considering in this case if we expect to run out of constant memory in our simulation. While limited to $\qty{64}{KB}$~\cite{cudaguide}, we expect all compile time matrices to fit in constant memory since their size only depends on the local degrees of freedom of the discretization.

Another potential avenue to accelerate memory accesses is to carefully prepare the layout of data. For the C++ CPU code, variables such as $\DB{S^{11}}$ are stored in row-major order, meaning that coefficients belonging to the same cell are continuous in memory. This locality is beneficial both for effective cache usage and for vectorized memory accesses. 
On the GPU the most efficient way to access global memory is through \emph{coalesced} reads whereby neighboring threads access neighboring addresses. Since each thread processes one cell, this can be achieved by storing variables in column-major order. 
Nonetheless, as we can see in \cref{tab:optimization}, the switch to column-major storage order for all fields causes a measurable slowdown. An investigation with a profiler revealed that the change does improve the memory access patterns as intended. The number of excessive sectors loaded from global memory decrease from $59\%$ for the row-major version to just $2\%$. However, this potential advantage is negated by the cache. Excessive data loaded due to the strided access of the row-major version is still needed by subsequent computations. Since this data is held in the cache, later global memory accesses can be circumvented. Furthermore, the column-major version performs more instructions for index computations, making the code slower overall.

A third option to reduce global memory accesses is to trade off reads with more computations.
This is beneficial when the code is memory-bound, as is often the case on the GPU, especially with classical linear algebra~{\cite{Dublish2017}.
In our code, the I/O can be reduced by re-computing the inverse parametric map $\DG{M^{-1}}$, which depends only on the geometry of the mesh and compile time constants. 
In particular, each matrix has a size of $n_S \times n_G$ while each mesh cell is fully described by $4$ vertices with $2$ values each. So, disregarding constants, even for a small dG-degree such as $n_G = 3$, fewer reads are required if we compute the matrices on-the-fly. 
Upon closer inspection, we also find that when stored in column-major order, vertex reads are coalesced while reads to $\DG{M^{-1}}$ are not, due to the fact that $\DG{M^{-1}}$ is implemented as an array of matrices. Since the Eigen matrix type only deals with two dimensions, adjusting the storage order of $\DG{M^{-1}}$ to allow for coalesced accesses would be difficult. 
Testing on an A100 GPU, we find that the on-the-fly map computation indeed delivers a speedup of $13\%$ over the CUDA baseline.

\subsection{SYCL}
SYCL is an open standard for heterogeneous computing developed by the Khronos group. With its most recent release SYCL 2020, the standard proposes a high-level API extending C++17 that allows the same code to run on various devices such as CPUs, GPUs and FPGAs. There are currently two major implementations of the standard, both of which are open source and based on LLVM. Development of AdaptiveCPP~\cite{Alpay2023}, previously known as hipSYCL and OpenSYCL, is lead by Heidelberg University. While various backends are available, the focus is on NVIDIA and AMD GPUs. The other major implementation is Data Parallel C++ (DPC++), developed by Intel. DPC++ primarily targets Intel CPUs, GPUs and FPGAs.

SYCL builds on top of standard C++ to minimize the effort of adapting existing code.
However, the SYCL standard forbids recursion and function pointers in kernel code~\cite{syclstd}, both of which are used in Eigen's expression templates. 
Albeit these function calls should be entirely inlined in the compiled code, DPC++ does not allow one to compile the \nextsim\ code because of these issues. 
AdaptiveCPP requires more effort for setup but the tool chain compiles Eigen.
We therefore limit our investigations to AdaptiveCPP in the following. 

SYCL automates device memory management and movement between host and device but it needs to be declared which memory will be used in a kernel. 
In particular, a \emph{buffer} needs to be defined, pointing to already allocated memory on the host. Then, a \emph{command group} is created which collects all information needed to run a task in parallel. Inside the command group, \emph{accessors} allow us to explicitly describe which buffers need to be accessed and how, i.e. read or write. Once put in a \emph{queue}, the SYCL runtime uses these memory requirements as well as optional dependencies on other command groups to select the best suited memory region to perform needed memory transfers and to schedule the execution. Inside the command group, we can declare a parallel for-loop and construct Eigen maps analogous to CUDA with pointers provided by the accessors.

In principle, we can attempt the same optimizations as with the CUDA code. While shared memory did not help in the case of CUDA, it is still of interest to see how its introduction affects the SYCL implementation, since memory management works differently.
To access \emph{local memory} in SYCL, which is the name used for CUDA's shared memory, we have to declare a \emph{local\_accessor} in the command buffer. 
In addition, local memory only makes sense in the context of thread blocks, so we need to use a more complicated for-loop which makes thread blocks explicit. Unfortunately, such a construct is known to perform far worse on CPU than a simple loop and work on reducing this gap is an active area of research~\cite{Meyer2023}. Therefore, if the code is to be run both on CPU and GPU, local memory should be introduced only in code paths specialized for the GPU. For the code snippet under study, this additional effort was not considered worthwhile. Returning to \cref{tab:optimization} we see using shared memory makes the kernel moderately slower. On the other hand, computing $\DG{M^{-1}}$ on-the-fly leads to a more substantial relative speedup over the AdaptiveCPP baseline than the same optimization in CUDA.

\subsection{Kokkos}
Kokkos is another programming model to enable heterogeneous computing in modern C++, currently with support for CPU as well as NVIDIA and AMD GPUs. Kokkos is developed as part of the Exascale Computing Project by the US Department of Energy. The main difference to SYCL is that Kokkos is a library while SYCL needs to be integrated into the compiler. The library-based approach greatly simplifies deployment of projects using Kokkos but potentially limits optimizations and some features.

Kokkos consists of macros and wrappers that provide a unified API for the different backends with the final code being processed by the chosen compiler. Therefore, we can once again use the same code, knowing already that it works in CUDA, and we can expect few differences to the pure CUDA implementation. The primary mechanism to manage memory in Kokkos are \emph{Views}, which are basically a shared pointer to a multi-dimensional array. Commonly, both a device view and a mirrored host view are created to facilitate data transfers. For our use case, it is possible to create a view on already allocated memory with the \emph{Unmanaged} trait. However, unmanaged views do not play well together with the mirrored views concept in backend agnostic code, leading to unnecessary copies during the execution on the CPU. Since in general the device view needs its own buffer, copies between the mirrored views will be performed regardless of whether they already reside in the same memory space. These extra copies can be avoided with a special case for view creation on CPU and are therefore not a major problem for portability. Once properly setup, data is accessible in the kernel through the device view and we can use the underlying pointer to create an Eigen map in the same manner as in CUDA.


Possible code optimizations in Kokkos are similar to those available in SYCL. The L1-cache, called \emph{scratch memory} in Kokkos, can be accessed by specifying a \emph{TeamPolicy} with a thread block size instead of using a simple parallel for-loop. Here a nuisance of the library becomes apparent as the total scratch memory needed for a particular kernel has to be set manually. Furthermore, parallelism described with explicit thread blocks has the same downside as in SYCL, namely that it leads to strongly degraded CPU performance. In our tests, recorded in \cref{tab:optimization}, we find that usage of scratch memory introduces a small overhead in Kokkos. On-the-fly map computation is, once more, beneficial and it results in a large speedup of $35\%$.

\subsection{PyTorch}
PyTorch~\cite{Paszke_PyTorch_An_Imperative_2019} is one of the most popular libraries for machine learning~\cite{aoun2022}. 
It consists of a simple-to-use Python frontend library and a high-performance C++ backend that has a dedicated compiler to optimize code execution and maps execution for different hardware such as CPUs, GPUs, and TPUs. Fully support are CPUs, NVIDIA GPUs and AMD GPUs.

To make effective use of PyTorch and the optimizations it implements, we have to reformulate our computations in terms of large tensors. 
For this, we remove the main loop in \cref{line:mainloop} and treat the element dimension $N$ as the batch dimension of variable size. The matrix-vector products become matrix-matrix products and element-wise operations remain unchanged. Some care is necessary to perform the products with the per-element inverse maps, e.g. \cref{line:mapprod}. Since we have a third dimension in $\DG{M^{-1}}$, this is not a standard matrix-matrix product. However, we can map this operation to a batched matrix-matrix product (bmm) by appending a dimension of size 1 to the second argument and removing it again afterwards (squeezing in PyTorch terminology). Alternatively we can formulate this computation as an element-wise product by adding a dimension corresponding to $n_s$ to the second argument, followed by a sum over that dimension. The latter operation turns out to be 4 to 5 times faster across different backends, indicating that PyTorch is not tuned for our use case with matrices much smaller than what is common in machine learning workloads.

To integrate the PyTorch code into our C++ simulation we have multiple options. With minor syntactic changes compared to the Python version, we can implement the computations directly with PyTorch's C++ API. However, this is suboptimal since each operation is executed as a separate kernel with no kernel-fusion taking place resulting in many reads and writes of the same data. 
A second option is to define the computation as a PyTorch model in Python. This model can be exported as \emph{TorchScript} and loaded in C++. Part of the C++ runtime is a just-in-time compiler which attempts to optimize the model execution on repeated use. However, more recent efforts to accelerate models have been focused on \emph{TorchDynamo}, a compiler first released with PyTorch 2.0 in March 2023. While the front-end of TorchDynamo is written in Python, various backends are available, some of which can be used without the Python runtime. Most promising among those we tested is the built-in \emph{TorchInductor} which leverages the compiler Triton~\cite{Tillet2019} to produce highly optimized fused-matrix multiplications~\cite{torchinductor}.
In particular, PyTorch 2.2 allows TorchInductor to generate a C++ wrapper function for the entire model which can then be integrated into C++ code.

\begin{figure}
\ifbuild
\begin{tikzpicture}
	\begin{axis}[
		ybar,
		enlarge x limits=0.15,
		legend style={legend columns=2},
		ylabel={runtime [\unit{s}]},
		symbolic x coords={Torch,TorchScript,TorchInductor},
		xtick=data,
		width=0.45\textwidth,
		height=4cm,
		ymin=0.0,%
		ymax=10.0,
		ybar=5pt,
		nodes near coords,
		nodes near coords align={vertical},
		]
		\addplot[tolcyan!20!black,fill=tolblue] coordinates {(Torch, 7.58) (TorchScript, 5.48) (TorchInductor, 5.79)};
		\addplot[tolcyan!20!black,fill=tolcyan] coordinates {(Torch,5.11) (TorchScript,3.64) (TorchInductor, 2.44)};
		\legend{bmm\\ $*$, sum\\}
	\end{axis}
\end{tikzpicture}
\else
\includegraphics{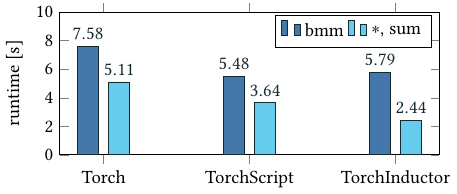}
\fi
\caption{Total time spend on the stress computation for the different PyTorch variants on an A100. The products with $\DG{M^{-1}}$ are implemented either as batched matrix-matrix product (bmm) or element-wise product and sum ($*$,sum).}
\label{plot:torch}
\end{figure}
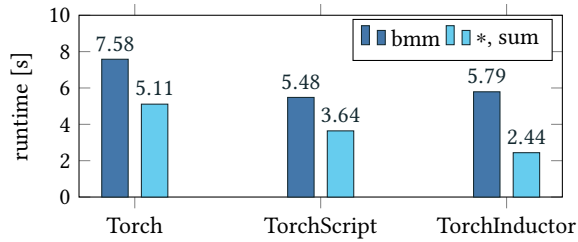

We compare the three proposed variants to integrate the PyTorch model into C++ in \cref{plot:torch}.
Albeit they use the same tensor primitives, the native C++ interface is considerably slower than TorchScript. The new compiler, TorchInductor, with its Triton optimized kernels, is significantly faster than the alternatives when using the element-wise product. When implemented with bmm, the compiler fails to optimize the operation due to a lack of GPU memory. Given that a batch size of \num{2.6d5} is already too large, while the operands require less than \qty{5}{MB} of memory, this points to it being an edge case not properly considered by the optimizer, instead of an actual memory limitation.

\subsection{Development and Deployment}
The development of dedicated CUDA code is time-consuming and error-prone. One purpose of the alternatives we considered in this work is to reduce this high development effort.
We therefore do a qualitative comparison of the different approaches considering ease of development but also deployment of the finished code on a target system.

With their modern C++ interface, both Kokkos and SYCL make it easier to write correct code compared to CUDA. Simplified resource management and stricter types prevent memory related bugs and make more errors visible at compile time. Using the simple parallelism constructs, a developer is not exposed to GPU specific scheduling based on blocks and grids. A further advantage of SYCL is that explicit annotations of device functions are unnecessary. Furthermore, SYCL's memory model fully automates transfers between host and device, eliminating another source of errors. It should be noted, however, that the more advanced C++ features used by Kokkos and SYCL can make the frameworks less approachable for non-C++ experts than the C-like interface of CUDA. PyTorch follows a completely different programming paradigm from the other options. While it takes time getting used to it when coming from system programming language background, development in PyTorch is overall much simpler. There is no memory management or explicit parallelism to take care of and rapid prototyping in Python is possible. A potential downside can be that some computations are hard to express in terms of tensor operations, in which case a low-level, manual implementation is still needed. Another downside is that code has to be completely rewritten in PyTorch, while for the options the C++ code can be largely reused.

When it comes to running the code on a target system, pure CUDA is the easiest. Usually pre-installed on clusters, no additional setup is required. Furthermore, CUDA (or Rocm) is a prerequisite for the other frameworks to use the GPU, so if a manual installation is needed, this effort is unavoidable for every framework. The Kokkos library can be easily integrated into a project's CMake based build system and works right out of the box. In combination with automatic fetching, e.g. via git submodule, the library setup becomes transparent. SYCL requires a specialized toolchain. For AdaptiveCPP, this means that it's compiler wrapper has to be build first from source. The manual configuration that is needed for AdaptiveCPP to find the proper compilers is cumbersome and might necessitate building a recent version of LLVM first. For PyTorch, prebuilds of the C++ library are available for all supported platforms. To use TorchInductor, the Python package is needed as well to generate the code on the target system, but it is easily acquired through a package manager. 

\section{Numerical Experiments}
\label{sec:experiments}

To analyze the performance of our implementations, we use the established VP benchmark of a moving cyclone over a sea ice region~\cite{Mehlmann2021}. 
%
The original C++ CPU version of the code has already been validated on this benchmark, see~\cite{Richter2023}. We can therefore ensure the correctness of our implementations by comparing to the CPU version. For time measurements, we simulate a duration of \qty{1}{\hour}, which requires $30$ advection steps and $3,000$ stress updates.

In the current GPU implementations, significant time is required to transfer memory between host and device.
Nonetheless, we focus on the kernel execution times in the following since the final objective of our work is a full GPU implementation of the dynamical core.
While transfers are still necessary for coupling with other models, the major effort of simulating the sea-ice dynamics is in the mEVP iteration. 
To ensure accurate timings, synchronization barriers are inserted as needed before and after the kernel invocation. In SYCL memory transfers are implicit, so we rely on the built-in profiling instead. Experiments are conducted in the following environment:
\begin{itemize}
	\item $2 \times$ AMD EPYC Rome 7402 CPU, $2 \times 24$ cores @ \qty{2.8}{GHz}
	\item NVIDIA A100 GPU, 40 GB HBM2e
	\item GCC-12.3
	\item CUDA 12.2
	\item Kokkos 4.1.0
	\item AdaptiveCPP 23.10.0 based on Clang-17.04
	\item PyTorch 2.2 Nightly (24-November-2023)
\end{itemize}


Of particular importance for coupled climate simulations is the performance scaling as a function of grid resolution and hence problem size.
With a fixed domain size of \qty{512}{km}, we reduce the resolution from \qty{4}{km} to \qty{0.25}{km}, which corresponds to an increase in the number of elements from \num{1.6d4} to \num{1.7d7} (i.e. one has a quadratic scaling in the resolution).
In \cref{plot:performance} we compare the best implementation for each approach as a function of elements.

\begin{figure}
\ifbuild
	\begin{tikzpicture}
		\begin{loglogaxis}[
			xlabel={\#elements},
			ylabel={runtime [\unit{s}]},
			width=0.499\textwidth,
			legend columns = 2,
			legend style={at={(0.5, 1.02)},anchor=south},
			grid=major,
			]
			\addplot table[x index=0,y index=2] {scaling.txt};
			\pgfplotsset{cycle list shift=-1}
			\addplot+[dashed] table[x index=0,y index=9] {scaling.txt};
			\addplot table[x index=0,y index=5] {scaling.txt};
			\addplot table[x index=0,y index=7] {scaling.txt};
			\addplot table[x index=0,y index=3] {scaling.txt};
			\addplot table[x index=0,y index=6] {scaling.txt};
			\addplot table[x index=0,y index=4] {scaling.txt};
			\pgfplotsset{cycle list shift=-2}
			\addplot+[dashed] table[x index=0,y index=6] {scaling_torch.txt};
			\addplot table[x index=0,y index=1] {scaling.txt};
			\legend{CUDA, CUDA (F32), AdaptiveCPP, AdaptiveCPP (CPU), Kokkos, Kokkos (CPU), TorchInductor, TorchInductor (TF32), OpenMP (CPU)}
		\end{loglogaxis}
	\end{tikzpicture}
\else
	\includegraphics{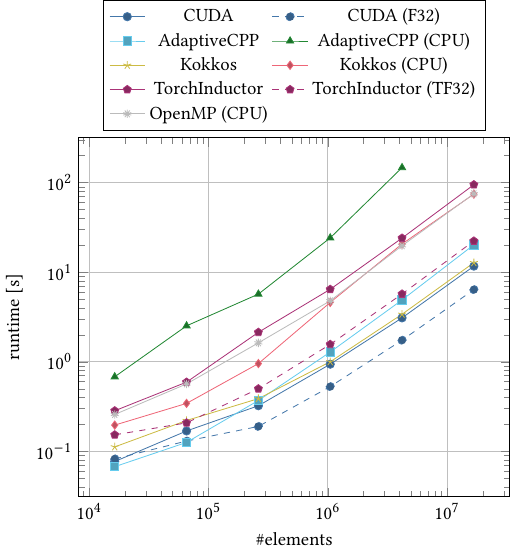}
\fi
	\caption{Timings of the stress update using the best performing version for each framework. The size of the mesh cells size is scaled from \qty{4}{km} to \qty{0.25}{km} while keeping the domain size constant to increase the number of elements. Dashed variants are run with lower floating-point precision.}
	\label{plot:performance}
\end{figure}
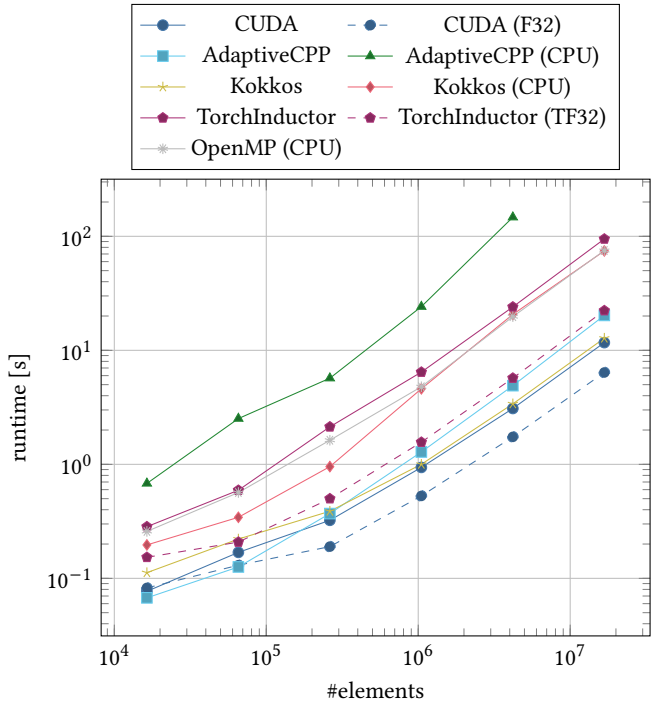

For CUDA, Kokkos and AdaptiveCPP, we compute the inverse maps on-the-fly. In case of Kokkos and AdaptiveCPP, the simple for-loop is used to run the update in parallel, which also makes the switch to CPU viable. For PyTorch, we take the implementation generated by TorchInductor.

Our CUDA implementation delivers a significant speedup over the OpenMP CPU reference implementation. For the smallest problem with \num{1.6d4} elements, CUDA is $3$ times faster, scaling up to $6.4$ for \num{1.7d7} elements. Kokkos asymptotically performs the same as CUDA on GPU and as OpenMP on CPU. This is to be expected since the very same compilers (NVCC and GCC) are used by Kokkos and only memory buffers and kernel dispatch are abstracted.
On small problems, Kokkos overhead makes it $50\%$ slower than CUDA but, surprisingly, the CPU version is slightly faster than raw OpenMP for the same number of elements. AdaptiveCPP scales worse than the other GPU accelerated codes, being $70\%$ slower than CUDA for the largest problem tested.
However, it still provides a significant improvement over the CPU OpenMP version. The good performance for small numbers is likely just an artifact from the differing time measuring method. On the CPU side, we could not get AdaptiveCPP to run properly. Best performance was achieved with a restriction to just 24 threads, indicating that the available CPUs are not utilized properly. 
The documentation states that performance of the CPU backend should be similar to raw OpenMP and that a significant deviation is likely caused by an improperly configured toolchain. However, we were unable to obtain stable results on three different systems, illustrating a greater difficulty in using the framework.
TorchInductor on GPU is slower than the OpenMP CPU code for every problem size tested and TorchInductor generated CPU code (not shown) is an order of magnitude slower than the GPU version.

The stress update can also be performed in single precision (F32). In our experiments this causes little degradation in the quality of the results. For CUDA, switching from double to float gives a speedup of $80\%$, which is in line with difference in peak performance of the A100 (double \qty{9.7}{TFLOPS}, float \qty{19.5} {TFLOPS}). We expect to see similar gains from using float in Kokkos and AdaptiveCPP. For TorchInductor, the relative speedup with float over the double version is much higher at $315\%$. As machine learning tasks rarely require double precision, the optimizer is likely tuned much more for single precision. The variant TorchInductor (TF32) from \cref{plot:performance}, additionally makes use of tensor cores which use a specialized tensor float format (TF32) that offers much higher peak performance for the A100 ($\qty{156}{TFLOPS}$). The actual speedup of $412\%$ over double is not quite as large, since not all operations can make use of the tensor cores. Furthermore, where it is possible, the matrices involved are too small to take full advantage of the tensor cores. In absolute terms, TorchInductor lacks behind the other options even with access to tensor cores and is therefore not suitable for our use case.

To test the portability of the heterogeneous compute frameworks we also run the experiments on a different system, equipped with an AMD MI250X GPU. All three, Kokkos, AdaptiveCPP and PyTorch work without modifications but only utilize half of the MI250X, since it is a dual chip design. AdaptiveCPP performs better on the AMD GPU, while Kokkos is somewhat slower than on the NVIDIA A100. Both thereby achieve a similar runtime, roughly $25\%$ higher than that of CUDA running on the A100. For PyTorch, the MI250X is $50\%$ slower than the A100 in our experiments.

\section{Conclusion}
\label{sec:conclusion}

We implemented and analyzed different options for the GPU parallelization of the \nextsim\ dynamical core.
Our results show that CUDA remains the most reliable option both in terms of performance and the toolchain. 
With the CUDA support of Eigen, we were also able to use the CPU C++ code essentially without modifications in CUDA.

Kokkos benefits the same from this library support, while SYCL does not need explicit support which makes it well suited for an incremental port of existing C++ code in general.
The streamlined memory model and simplified parallel constructs of Kokkos and SYCL facilitate more effective development, but at some performance cost.
One remaining issue is that using dedicated GPU features such as shared memory leads to code that is very inefficient on the CPU, breaking the promise of the heterogeneous computing paradigm.
However, our study demonstrates that this specialization is not always needed to achieve good performance and we can therefore recommend Kokkos as an alternative to CUDA.
While SYCL shares the same benefits on paper, it suffers from immature implementations and is currently too unreliable.
PyTorch currently lacks far behind the more conventional options in terms of performance and is therefore mostly worth considering for rapid prototyping. However, the underlying compilers are developing quickly and in our case, the optimizer heuristics are clearly far from optimal yet, so we expect further improvements in the future. Furthermore, PyTorch and similar tools are worth keeping an eye on, both for there ease of development and the access to automatic differentiation they provide. The latter is of great relevance for hybrid methods that combine a conventional discretization with a machine learning component, e.g.~\cite{Bedrunka2021, Kochkov2021, Demeure2023, Kochkov2023}.

For the full port of \nextsim\ we choose Kokkos, because it is well maintained and demonstrates good performance across GPUs and CPUs. For the complete simulation running in double precision on GPU, we expect a speedup of factor $4$ or, alternatively, a doubling of the resolution with the same runtime. Since \nextsim\ is still in active development, performance comparisons with currently used models are difficult. However, to move from the current \qty{10}{km} resolution sea ice models~\cite{Olason2021, Hutter2022} to practical kilometer-scale models more work will likely be needed. In future work, we thus want to investigate the effects of mixed precision on the simulation in greater depth.



\section{Code Availability}
The project \nextsim\ is under active development and hosted on GitHub~(\url{https://github.com/nextsimhub/nextsimdg}). The self-contained repository with the GPU implementations and experiments described in this paper is available at  \url{https://kosinus.math.uni-magdeburg.de/Thanduriel/dynamical_core}.

\section{Acknowledgments}
This project is supported by Schmidt Futures – a philanthropic initiative that seeks to improve societal outcomes through the development of emerging science and technologies.

\bibliographystyle{ACM-Reference-Format}
\bibliography{../library}


\begin{thebibliography}{38}


\ifx \showCODEN    \undefined \def \showCODEN     #1{\unskip}     \fi
\ifx \showDOI      \undefined \def \showDOI       #1{#1}\fi
\ifx \showISBNx    \undefined \def \showISBNx     #1{\unskip}     \fi
\ifx \showISBNxiii \undefined \def \showISBNxiii  #1{\unskip}     \fi
\ifx \showISSN     \undefined \def \showISSN      #1{\unskip}     \fi
\ifx \showLCCN     \undefined \def \showLCCN      #1{\unskip}     \fi
\ifx \shownote     \undefined \def \shownote      #1{#1}          \fi
\ifx \showarticletitle \undefined \def \showarticletitle #1{#1}   \fi
\ifx \showURL      \undefined \def \showURL       {\relax}        \fi
\providecommand\bibfield[2]{#2}
\providecommand\bibinfo[2]{#2}
\providecommand\natexlab[1]{#1}
\providecommand\showeprint[2][]{arXiv:#2}

\bibitem[xla(2023)]%
        {xlarepo}
 \bibinfo{year}{2023}\natexlab{}.
\newblock \bibinfo{title}{{XLA: an open-source machine learning compiler}}.
\newblock
\newblock
\urldef\tempurl%
\url{https://github.com/openxla/xla}
\showURL{%
\tempurl}
\newblock
\shownote{[accessed 28-November-2023]}.


\bibitem[Alpay and Heuveline(2023)]%
        {Alpay2023}
\bibfield{author}{\bibinfo{person}{Aksel Alpay} {and} \bibinfo{person}{Vincent
  Heuveline}.} \bibinfo{year}{2023}\natexlab{}.
\newblock \showarticletitle{One Pass to Bind Them: The First Single-Pass {SYCL}
  Compiler with Unified Code Representation Across Backends}. In
  \bibinfo{booktitle}{\emph{International Workshop on {OpenCL}}}.
  \bibinfo{publisher}{{ACM}}.
\newblock
\urldef\tempurl%
\url{https://doi.org/10.1145/3585341.3585351}
\showDOI{\tempurl}


\bibitem[aoun et~al\mbox{.}(2022)]%
        {aoun2022}
\bibfield{author}{\bibinfo{person}{Mohamed Raed~El aoun},
  \bibinfo{person}{Lionel~Nganyewou Tidjon}, \bibinfo{person}{Ben Rombaut},
  \bibinfo{person}{Foutse Khomh}, {and} \bibinfo{person}{Ahmed~E. Hassan}.}
  \bibinfo{year}{2022}\natexlab{}.
\newblock \bibinfo{title}{An Empirical Study of Library Usage and Dependency in
  Deep Learning Frameworks}.
\newblock
\newblock
\urldef\tempurl%
\url{https://doi.org/10.48550/ARXIV.2211.15733}
\showDOI{\tempurl}


\bibitem[Bauer et~al\mbox{.}(2021a)]%
        {Bauer2021a}
\bibfield{author}{\bibinfo{person}{Peter Bauer}, \bibinfo{person}{Peter~D.
  Dueben}, \bibinfo{person}{Torsten Hoefler}, \bibinfo{person}{Tiago Quintino},
  \bibinfo{person}{Thomas~C. Schulthess}, {and} \bibinfo{person}{Nils~P.
  Wedi}.} \bibinfo{year}{2021}\natexlab{a}.
\newblock \showarticletitle{The digital revolution of Earth-system science}.
\newblock \bibinfo{journal}{\emph{Nature Computational Science}}
  \bibinfo{volume}{1}, \bibinfo{number}{2} (\bibinfo{date}{Feb.}
  \bibinfo{year}{2021}), \bibinfo{pages}{104--113}.
\newblock
\showISSN{2662-8457}
\urldef\tempurl%
\url{https://doi.org/10.1038/s43588-021-00023-0}
\showDOI{\tempurl}


\bibitem[Bauer et~al\mbox{.}(2021b)]%
        {Bauer2021}
\bibfield{author}{\bibinfo{person}{Peter Bauer}, \bibinfo{person}{Bjorn
  Stevens}, {and} \bibinfo{person}{Wilco Hazeleger}.}
  \bibinfo{year}{2021}\natexlab{b}.
\newblock \showarticletitle{A digital twin of Earth for the green transition}.
\newblock \bibinfo{journal}{\emph{Nature Climate Change}} \bibinfo{volume}{11},
  \bibinfo{number}{2} (\bibinfo{year}{2021}), \bibinfo{pages}{80--83}.
\newblock
\showISBNx{1758-6798}
\urldef\tempurl%
\url{https://doi.org/10.1038/s41558-021-00986-y}
\showDOI{\tempurl}


\bibitem[Bedrunka et~al\mbox{.}(2021)]%
        {Bedrunka2021}
\bibfield{author}{\bibinfo{person}{Mario~Christopher Bedrunka},
  \bibinfo{person}{Dominik Wilde}, \bibinfo{person}{Martin Kliemank},
  \bibinfo{person}{Dirk Reith}, \bibinfo{person}{Holger Foysi}, {and}
  \bibinfo{person}{Andreas Krämer}.} \bibinfo{year}{2021}\natexlab{}.
\newblock \bibinfo{booktitle}{\emph{Lettuce: PyTorch-Based Lattice Boltzmann
  Framework}}.
\newblock \bibinfo{publisher}{Springer International Publishing},
  \bibinfo{pages}{40--55}.
\newblock
\showISBNx{9783030905392}
\showISSN{1611-3349}
\urldef\tempurl%
\url{https://doi.org/10.1007/978-3-030-90539-2_3}
\showDOI{\tempurl}


\bibitem[Bouillon et~al\mbox{.}(2013)]%
        {Bouillon2013}
\bibfield{author}{\bibinfo{person}{Sylvain Bouillon}, \bibinfo{person}{Thierry
  Fichefet}, \bibinfo{person}{Vincent Legat}, {and} \bibinfo{person}{Gurvan
  Madec}.} \bibinfo{year}{2013}\natexlab{}.
\newblock \showarticletitle{The elastic{\textendash}viscous{\textendash}plastic
  method revisited}.
\newblock \bibinfo{journal}{\emph{Ocean Modelling}}  \bibinfo{volume}{71}
  (\bibinfo{date}{nov} \bibinfo{year}{2013}), \bibinfo{pages}{2--12}.
\newblock
\urldef\tempurl%
\url{https://doi.org/10.1016/j.ocemod.2013.05.013}
\showDOI{\tempurl}


\bibitem[Bradbury et~al\mbox{.}(2018)]%
        {jax2018github}
\bibfield{author}{\bibinfo{person}{James Bradbury}, \bibinfo{person}{Roy
  Frostig}, \bibinfo{person}{Peter Hawkins}, \bibinfo{person}{Matthew~James
  Johnson}, \bibinfo{person}{Chris Leary}, \bibinfo{person}{Dougal Maclaurin},
  \bibinfo{person}{George Necula}, \bibinfo{person}{Adam Paszke},
  \bibinfo{person}{Jake Vander{P}las}, \bibinfo{person}{Skye
  Wanderman-{M}ilne}, {and} \bibinfo{person}{Qiao Zhang}.}
  \bibinfo{year}{2018}\natexlab{}.
\newblock \bibinfo{booktitle}{\emph{{JAX}: composable transformations of
  {P}ython+{N}um{P}y programs}}.
\newblock
\urldef\tempurl%
\url{http://github.com/google/jax}
\showURL{%
\tempurl}


\bibitem[Cao et~al\mbox{.}(2023)]%
        {Cao2023}
\bibfield{author}{\bibinfo{person}{Kai Cao}, \bibinfo{person}{Qizhong Wu},
  \bibinfo{person}{Lingling Wang}, \bibinfo{person}{Nan Wang},
  \bibinfo{person}{Huaqiong Cheng}, \bibinfo{person}{Xiao Tang},
  \bibinfo{person}{Dongqing Li}, {and} \bibinfo{person}{Lanning Wang}.}
  \bibinfo{year}{2023}\natexlab{}.
\newblock \showarticletitle{{GPU}-{HADVPPM} V1.0: a high-efficiency parallel
  {GPU} design of the piecewise parabolic method ({PPM}) for horizontal
  advection in an air quality model ({CAMx} V6.10)}.
\newblock \bibinfo{journal}{\emph{Geoscientific Model Development}}
  \bibinfo{volume}{16}, \bibinfo{number}{15} (\bibinfo{date}{aug}
  \bibinfo{year}{2023}), \bibinfo{pages}{4367--4383}.
\newblock
\urldef\tempurl%
\url{https://doi.org/10.5194/gmd-16-4367-2023}
\showDOI{\tempurl}


\bibitem[Demeure et~al\mbox{.}(2023)]%
        {Demeure2023}
\bibfield{author}{\bibinfo{person}{Nestor Demeure}, \bibinfo{person}{Theodore
  Kisner}, \bibinfo{person}{Reijo Keskitalo}, \bibinfo{person}{Rollin Thomas},
  \bibinfo{person}{Julian Borrill}, {and} \bibinfo{person}{Wahid Bhimji}.}
  \bibinfo{year}{2023}\natexlab{}.
\newblock \showarticletitle{High-level GPU code: a case study examining JAX and
  OpenMP.}. In \bibinfo{booktitle}{\emph{Proceedings of the SC ’23 Workshops
  of The International Conference on High Performance Computing, Network,
  Storage, and Analysis}} \emph{(\bibinfo{series}{SC-W 2023})}.
  \bibinfo{publisher}{ACM}.
\newblock
\urldef\tempurl%
\url{https://doi.org/10.1145/3624062.3624186}
\showDOI{\tempurl}


\bibitem[Dublish et~al\mbox{.}(2017)]%
        {Dublish2017}
\bibfield{author}{\bibinfo{person}{Saumay Dublish}, \bibinfo{person}{Vijay
  Nagarajan}, {and} \bibinfo{person}{Nigel Topham}.}
  \bibinfo{year}{2017}\natexlab{}.
\newblock \showarticletitle{Evaluating and mitigating bandwidth bottlenecks
  across the memory hierarchy in GPUs}. In \bibinfo{booktitle}{\emph{2017 IEEE
  International Symposium on Performance Analysis of Systems and Software
  (ISPASS)}}. \bibinfo{publisher}{IEEE}.
\newblock
\urldef\tempurl%
\url{https://doi.org/10.1109/ispass.2017.7975295}
\showDOI{\tempurl}


\bibitem[Feltham(2008)]%
        {Feltham2008}
\bibfield{author}{\bibinfo{person}{D.L. Feltham}.}
  \bibinfo{year}{2008}\natexlab{}.
\newblock \showarticletitle{Sea Ice Rheology}.
\newblock \bibinfo{journal}{\emph{Annual Review of Fluid Mechanics}}
  \bibinfo{volume}{40} (\bibinfo{year}{2008}), \bibinfo{pages}{91--112}.
\newblock
\urldef\tempurl%
\url{https://doi.org/10.1146/annurev.fluid.40.111406.102151}
\showDOI{\tempurl}


\bibitem[Guennebaud et~al\mbox{.}(2010)]%
        {eigenweb}
\bibfield{author}{\bibinfo{person}{Ga\"{e}l Guennebaud},
  \bibinfo{person}{Beno\^{i}t Jacob}, {et~al\mbox{.}}}
  \bibinfo{year}{2010}\natexlab{}.
\newblock \bibinfo{title}{Eigen v3}.
\newblock \bibinfo{howpublished}{http://eigen.tuxfamily.org}.
\newblock


\bibitem[Hibler(1979)]%
        {Hibler1979}
\bibfield{author}{\bibinfo{person}{W.~D. Hibler}.}
  \bibinfo{year}{1979}\natexlab{}.
\newblock \showarticletitle{A Dynamic Thermodynamic Sea Ice Model}.
\newblock \bibinfo{journal}{\emph{Journal of Physical Oceanography}}
  \bibinfo{volume}{9}, \bibinfo{number}{4} (\bibinfo{date}{jul}
  \bibinfo{year}{1979}), \bibinfo{pages}{815--846}.
\newblock
\urldef\tempurl%
\url{https://doi.org/10.1175/1520-0485(1979)009<0815:adtsim>2.0.co;2}
\showDOI{\tempurl}


\bibitem[Hutter et~al\mbox{.}(2022)]%
        {Hutter2022}
\bibfield{author}{\bibinfo{person}{Nils Hutter}, \bibinfo{person}{Amélie
  Bouchat}, \bibinfo{person}{Frédéric Dupont}, \bibinfo{person}{Dmitry
  Dukhovskoy}, \bibinfo{person}{Nikolay Koldunov}, \bibinfo{person}{Younjoo~J.
  Lee}, \bibinfo{person}{Jean‐François Lemieux}, \bibinfo{person}{Camille
  Lique}, \bibinfo{person}{Martin Losch}, \bibinfo{person}{Wieslaw Maslowski},
  \bibinfo{person}{Paul~G. Myers}, \bibinfo{person}{Einar Ólason},
  \bibinfo{person}{Pierre Rampal}, \bibinfo{person}{Till Rasmussen},
  \bibinfo{person}{Claude Talandier}, \bibinfo{person}{Bruno Tremblay}, {and}
  \bibinfo{person}{Qiang Wang}.} \bibinfo{year}{2022}\natexlab{}.
\newblock \showarticletitle{Sea Ice Rheology Experiment (SIREx): 2. Evaluating
  Linear Kinematic Features in High‐Resolution Sea Ice Simulations}.
\newblock \bibinfo{journal}{\emph{Journal of Geophysical Research: Oceans}}
  \bibinfo{volume}{127}, \bibinfo{number}{4} (\bibinfo{date}{April}
  \bibinfo{year}{2022}).
\newblock
\showISSN{2169-9291}
\urldef\tempurl%
\url{https://doi.org/10.1029/2021jc017666}
\showDOI{\tempurl}


\bibitem[Ikuyajolu et~al\mbox{.}(2023)]%
        {Ikuyajolu2023}
\bibfield{author}{\bibinfo{person}{Olawale~James Ikuyajolu},
  \bibinfo{person}{Luke~Van Roekel}, \bibinfo{person}{Steven~R. Brus},
  \bibinfo{person}{Erin~E. Thomas}, \bibinfo{person}{Yi Deng}, {and}
  \bibinfo{person}{Sarat Sreepathi}.} \bibinfo{year}{2023}\natexlab{}.
\newblock \showarticletitle{Porting the {WAVEWATCH} {III} (v6.07) wave action
  source terms to {GPU}}.
\newblock \bibinfo{journal}{\emph{Geoscientific Model Development}}
  \bibinfo{volume}{16}, \bibinfo{number}{4} (\bibinfo{date}{mar}
  \bibinfo{year}{2023}), \bibinfo{pages}{1445--1458}.
\newblock
\urldef\tempurl%
\url{https://doi.org/10.5194/gmd-16-1445-2023}
\showDOI{\tempurl}


\bibitem[Jakob et~al\mbox{.}(2023)]%
        {Jakob2023}
\bibfield{author}{\bibinfo{person}{Christian Jakob}, \bibinfo{person}{Andrew
  Gettelman}, {and} \bibinfo{person}{Andrew Pitman}.}
  \bibinfo{year}{2023}\natexlab{}.
\newblock \showarticletitle{The need to operationalize climate modelling}.
\newblock \bibinfo{journal}{\emph{Nature Climate Change}} \bibinfo{volume}{13},
  \bibinfo{number}{11} (\bibinfo{year}{2023}), \bibinfo{pages}{1158--1160}.
\newblock
\showISBNx{1758-6798}
\urldef\tempurl%
\url{https://doi.org/10.1038/s41558-023-01849-4}
\showDOI{\tempurl}


\bibitem[Kochkov et~al\mbox{.}(2021)]%
        {Kochkov2021}
\bibfield{author}{\bibinfo{person}{Dmitrii Kochkov}, \bibinfo{person}{Jamie~A.
  Smith}, \bibinfo{person}{Ayya Alieva}, \bibinfo{person}{Qing Wang},
  \bibinfo{person}{Michael~P. Brenner}, {and} \bibinfo{person}{Stephan Hoyer}.}
  \bibinfo{year}{2021}\natexlab{}.
\newblock \showarticletitle{Machine learning–accelerated computational fluid
  dynamics}.
\newblock \bibinfo{journal}{\emph{Proceedings of the National Academy of
  Sciences}} \bibinfo{volume}{118}, \bibinfo{number}{21} (\bibinfo{date}{May}
  \bibinfo{year}{2021}).
\newblock
\showISSN{1091-6490}
\urldef\tempurl%
\url{https://doi.org/10.1073/pnas.2101784118}
\showDOI{\tempurl}


\bibitem[Kochkov et~al\mbox{.}(2023)]%
        {Kochkov2023}
\bibfield{author}{\bibinfo{person}{Dmitrii Kochkov}, \bibinfo{person}{Janni
  Yuval}, \bibinfo{person}{Ian Langmore}, \bibinfo{person}{Peter Norgaard},
  \bibinfo{person}{Jamie Smith}, \bibinfo{person}{Griffin Mooers},
  \bibinfo{person}{James Lottes}, \bibinfo{person}{Stephan Rasp},
  \bibinfo{person}{Peter D{\"u}ben}, \bibinfo{person}{Milan Kl{\"o}wer},
  \bibinfo{person}{Sam Hatfield}, \bibinfo{person}{Peter Battaglia},
  \bibinfo{person}{Alvaro Sanchez-Gonzalez}, \bibinfo{person}{Matthew Willson},
  \bibinfo{person}{Michael~P. Brenner}, {and} \bibinfo{person}{Stephan Hoyer}.}
  \bibinfo{year}{2023}\natexlab{}.
\newblock \bibinfo{title}{Neural General Circulation Models}.
\newblock
\newblock
\showeprint[arxiv]{2311.07222}~[physics.ao-ph]


\bibitem[Mehlmann et~al\mbox{.}(2021)]%
        {Mehlmann2021}
\bibfield{author}{\bibinfo{person}{C. Mehlmann}, \bibinfo{person}{S. Danilov},
  \bibinfo{person}{M. Losch}, \bibinfo{person}{J.~F. Lemieux},
  \bibinfo{person}{N. Hutter}, \bibinfo{person}{T. Richter},
  \bibinfo{person}{P. Blain}, \bibinfo{person}{E.~C. Hunke}, {and}
  \bibinfo{person}{P. Korn}.} \bibinfo{year}{2021}\natexlab{}.
\newblock \showarticletitle{Simulating Linear Kinematic Features in
  Viscous‐Plastic Sea Ice Models on Quadrilateral and Triangular Grids With
  Different Variable Staggering}.
\newblock \bibinfo{journal}{\emph{Journal of Advances in Modeling Earth
  Systems}} \bibinfo{volume}{13}, \bibinfo{number}{11} (\bibinfo{date}{Oct.}
  \bibinfo{year}{2021}).
\newblock
\showISSN{1942-2466}
\urldef\tempurl%
\url{https://doi.org/10.1029/2021ms002523}
\showDOI{\tempurl}


\bibitem[Mehlmann and Richter(2017)]%
        {MehlmannRichter2017}
\bibfield{author}{\bibinfo{person}{C. Mehlmann} {and} \bibinfo{person}{T.
  Richter}.} \bibinfo{year}{2017}\natexlab{}.
\newblock \showarticletitle{A modified global Newton solver for viscous-plastic
  sea ice models}.
\newblock \bibinfo{journal}{\emph{Ocean Modeling}}  \bibinfo{volume}{116}
  (\bibinfo{year}{2017}), \bibinfo{pages}{96--107}.
\newblock
\urldef\tempurl%
\url{https://doi.org/10.1016/j.ocemod.2017.06.001}
\showDOI{\tempurl}


\bibitem[Meyer et~al\mbox{.}(2023)]%
        {Meyer2023}
\bibfield{author}{\bibinfo{person}{Joachim Meyer}, \bibinfo{person}{Aksel
  Alpay}, \bibinfo{person}{Sebastian Hack}, \bibinfo{person}{Holger Fröning},
  {and} \bibinfo{person}{Vincent Heuveline}.} \bibinfo{year}{2023}\natexlab{}.
\newblock \showarticletitle{Implementation Techniques for SPMD Kernels on
  CPUs}. In \bibinfo{booktitle}{\emph{International Workshop on OpenCL}}
  \emph{(\bibinfo{series}{IWOCL ’23})}. \bibinfo{publisher}{ACM}.
\newblock
\urldef\tempurl%
\url{https://doi.org/10.1145/3585341.3585342}
\showDOI{\tempurl}


\bibitem[NVIDIA(2023a)]%
        {cudaguide}
\bibfield{author}{\bibinfo{person}{NVIDIA}.} \bibinfo{year}{2023}\natexlab{a}.
\newblock \bibinfo{title}{{CUDA C++ Programming Guide}}.
\newblock
\newblock
\urldef\tempurl%
\url{https://docs.nvidia.com/cuda/cuda-c-programming-guide/index.html}
\showURL{%
\tempurl}
\newblock
\shownote{[accessed 28-November-2023]}.


\bibitem[NVIDIA(2023b)]%
        {tensorrtmain}
\bibfield{author}{\bibinfo{person}{NVIDIA}.} \bibinfo{year}{2023}\natexlab{b}.
\newblock \bibinfo{title}{{TensorRT: an SDK for high-performance deep learning
  inference}}.
\newblock
\newblock
\urldef\tempurl%
\url{https://developer.nvidia.com/tensorrt}
\showURL{%
\tempurl}
\newblock
\shownote{[accessed 28-November-2023]}.


\bibitem[Paszke et~al\mbox{.}(2019)]%
        {Paszke_PyTorch_An_Imperative_2019}
\bibfield{author}{\bibinfo{person}{Adam Paszke}, \bibinfo{person}{Sam Gross},
  \bibinfo{person}{Francisco Massa}, \bibinfo{person}{Adam Lerer},
  \bibinfo{person}{James Bradbury}, \bibinfo{person}{Gregory Chanan},
  \bibinfo{person}{Trevor Killeen}, \bibinfo{person}{Zeming Lin},
  \bibinfo{person}{Natalia Gimelshein}, \bibinfo{person}{Luca Antiga},
  \bibinfo{person}{Alban Desmaison}, \bibinfo{person}{Andreas Kopf},
  \bibinfo{person}{Edward Yang}, \bibinfo{person}{Zachary DeVito},
  \bibinfo{person}{Martin Raison}, \bibinfo{person}{Alykhan Tejani},
  \bibinfo{person}{Sasank Chilamkurthy}, \bibinfo{person}{Benoit Steiner},
  \bibinfo{person}{Lu Fang}, \bibinfo{person}{Junjie Bai}, {and}
  \bibinfo{person}{Soumith Chintala}.} \bibinfo{year}{2019}\natexlab{}.
\newblock \showarticletitle{{PyTorch: An Imperative Style, High-Performance
  Deep Learning Library}}. In \bibinfo{booktitle}{\emph{Advances in Neural
  Information Processing Systems 32}},
  \bibfield{editor}{\bibinfo{person}{H.~Wallach},
  \bibinfo{person}{H.~Larochelle}, \bibinfo{person}{A.~Beygelzimer},
  \bibinfo{person}{F.~d'Alché Buc}, \bibinfo{person}{E.~Fox}, {and}
  \bibinfo{person}{R.~Garnett}} (Eds.). \bibinfo{publisher}{Curran Associates,
  Inc.}, \bibinfo{pages}{8024--8035}.
\newblock
\urldef\tempurl%
\url{http://papers.neurips.cc/paper/9015-pytorch-an-imperative-style-high-performance-deep-learning-library.pdf}
\showURL{%
\tempurl}


\bibitem[Richter et~al\mbox{.}(2023)]%
        {Richter2023}
\bibfield{author}{\bibinfo{person}{Thomas Richter},
  \bibinfo{person}{V{\'{e}}ronique Dansereau}, \bibinfo{person}{Christian
  Lessig}, {and} \bibinfo{person}{Piotr Minakowski}.}
  \bibinfo{year}{2023}\natexlab{}.
\newblock \showarticletitle{A dynamical core based on a discontinuous Galerkin
  method for higher-order finite-element sea ice modeling}.
\newblock \bibinfo{journal}{\emph{Geoscientific Model Development}}
  \bibinfo{volume}{16}, \bibinfo{number}{13} (\bibinfo{date}{jul}
  \bibinfo{year}{2023}), \bibinfo{pages}{3907--3926}.
\newblock
\urldef\tempurl%
\url{https://doi.org/10.5194/gmd-16-3907-2023}
\showDOI{\tempurl}


\bibitem[Sauer and Mu{\~{n}}oz-Esparza(2020)]%
        {Sauer2020}
\bibfield{author}{\bibinfo{person}{Jeremy~A. Sauer} {and}
  \bibinfo{person}{Domingo Mu{\~{n}}oz-Esparza}.}
  \bibinfo{year}{2020}\natexlab{}.
\newblock \showarticletitle{The {FastEddy}{\textregistered} Resident-{GPU}
  Accelerated Large-Eddy Simulation Framework: Model Formulation,
  Dynamical-Core Validation and Performance Benchmarks}.
\newblock \bibinfo{journal}{\emph{Journal of Advances in Modeling Earth
  Systems}} \bibinfo{volume}{12}, \bibinfo{number}{11} (\bibinfo{date}{nov}
  \bibinfo{year}{2020}).
\newblock
\urldef\tempurl%
\url{https://doi.org/10.1029/2020ms002100}
\showDOI{\tempurl}


\bibitem[Schär et~al\mbox{.}(2020)]%
        {Schaer2020}
\bibfield{author}{\bibinfo{person}{Christoph Schär}, \bibinfo{person}{Oliver
  Fuhrer}, \bibinfo{person}{Andrea Arteaga}, \bibinfo{person}{Nikolina Ban},
  \bibinfo{person}{Christophe Charpilloz}, \bibinfo{person}{Salvatore~Di
  Girolamo}, \bibinfo{person}{Laureline Hentgen}, \bibinfo{person}{Torsten
  Hoefler}, \bibinfo{person}{Xavier Lapillonne}, \bibinfo{person}{David
  Leutwyler}, \bibinfo{person}{Katherine Osterried}, \bibinfo{person}{Davide
  Panosetti}, \bibinfo{person}{Stefan Rüdisühli}, \bibinfo{person}{Linda
  Schlemmer}, \bibinfo{person}{Thomas~C. Schulthess}, \bibinfo{person}{Michael
  Sprenger}, \bibinfo{person}{Stefano Ubbiali}, {and} \bibinfo{person}{Heini
  Wernli}.} \bibinfo{year}{2020}\natexlab{}.
\newblock \showarticletitle{Kilometer-Scale Climate Models: Prospects and
  Challenges}.
\newblock \bibinfo{journal}{\emph{Bulletin of the American Meteorological
  Society}} \bibinfo{volume}{101}, \bibinfo{number}{5} (\bibinfo{date}{may}
  \bibinfo{year}{2020}), \bibinfo{pages}{E567--E587}.
\newblock
\urldef\tempurl%
\url{https://doi.org/10.1175/bams-d-18-0167.1}
\showDOI{\tempurl}


\bibitem[Stevens et~al\mbox{.}(2019)]%
        {Stevens2019}
\bibfield{author}{\bibinfo{person}{Bjorn Stevens}, \bibinfo{person}{Masaki
  Satoh}, \bibinfo{person}{Ludovic Auger}, \bibinfo{person}{Joachim Biercamp},
  \bibinfo{person}{Christopher~S. Bretherton}, \bibinfo{person}{Xi Chen},
  \bibinfo{person}{Peter D{\"u}ben}, \bibinfo{person}{Falko Judt},
  \bibinfo{person}{Marat Khairoutdinov}, \bibinfo{person}{Daniel Klocke},
  \bibinfo{person}{Chihiro Kodama}, \bibinfo{person}{Luis Kornblueh},
  \bibinfo{person}{Shian-Jiann Lin}, \bibinfo{person}{Philipp Neumann},
  \bibinfo{person}{William~M. Putman}, \bibinfo{person}{Niklas R{\"o}ber},
  \bibinfo{person}{Ryosuke Shibuya}, \bibinfo{person}{Benoit Vanniere},
  \bibinfo{person}{Pier~Luigi Vidale}, \bibinfo{person}{Nils Wedi}, {and}
  \bibinfo{person}{Linjiong Zhou}.} \bibinfo{year}{2019}\natexlab{}.
\newblock \showarticletitle{DYAMOND: the DYnamics of the Atmospheric general
  circulation Modeled On Non-hydrostatic Domains}.
\newblock \bibinfo{journal}{\emph{Progress in Earth and Planetary Science}}
  \bibinfo{volume}{6}, \bibinfo{number}{1} (\bibinfo{year}{2019}),
  \bibinfo{pages}{61}.
\newblock
\showISBNx{2197-4284}
\urldef\tempurl%
\url{https://doi.org/10.1186/s40645-019-0304-z}
\showDOI{\tempurl}


\bibitem[Sun et~al\mbox{.}(2023)]%
        {Sun2023}
\bibfield{author}{\bibinfo{person}{Jian Sun}, \bibinfo{person}{John~M. Dennis},
  \bibinfo{person}{Sheri~A. Mickelson}, \bibinfo{person}{Brian Vanderwende},
  \bibinfo{person}{Andrew Gettelman}, {and} \bibinfo{person}{Katherine
  Thayer-Calder}.} \bibinfo{year}{2023}\natexlab{}.
\newblock \showarticletitle{Acceleration of the Parameterization of Unified
  Microphysics Across Scales ({PUMAS}) on the Graphics Processing Unit ({GPU})
  With Directive-Based Methods}.
\newblock \bibinfo{journal}{\emph{Journal of Advances in Modeling Earth
  Systems}} \bibinfo{volume}{15}, \bibinfo{number}{5} (\bibinfo{date}{may}
  \bibinfo{year}{2023}).
\newblock
\urldef\tempurl%
\url{https://doi.org/10.1029/2022ms003515}
\showDOI{\tempurl}


\bibitem[team(2023)]%
        {torchinductor}
\bibfield{author}{\bibinfo{person}{PyTorch team}.}
  \bibinfo{year}{2023}\natexlab{}.
\newblock \bibinfo{title}{{TorchInductor: a PyTorch-native Compiler}}.
\newblock
\newblock
\urldef\tempurl%
\url{https://dev-discuss.pytorch.org/t/torchinductor-a-pytorch-native-compiler-with-define-by-run-ir-and-symbolic-shapes/747}
\showURL{%
\tempurl}
\newblock
\shownote{[accessed 28-November-2023]}.


\bibitem[The Khronos~Group(2023a)]%
        {syclstd}
\bibfield{author}{\bibinfo{person}{Inc. The Khronos~Group}.}
  \bibinfo{year}{2023}\natexlab{a}.
\newblock \bibinfo{title}{{SYCL 2020 standard - language restrictions}}.
\newblock
\newblock
\urldef\tempurl%
\url{https://registry.khronos.org/SYCL/specs/sycl-2020/html/sycl-2020.html#_language_restrictions_in_kernels}
\showURL{%
\tempurl}
\newblock
\shownote{[accessed 21-November-2023]}.


\bibitem[The Khronos~Group(2023b)]%
        {syclmain}
\bibfield{author}{\bibinfo{person}{Inc. The Khronos~Group}.}
  \bibinfo{year}{2023}\natexlab{b}.
\newblock \bibinfo{title}{{SYCL: a cross-platform abstraction layer for
  heterogeneous computing}}.
\newblock
\newblock
\urldef\tempurl%
\url{https://www.khronos.org/sycl}
\showURL{%
\tempurl}
\newblock
\shownote{[accessed 30-November-2023]}.


\bibitem[Tillet et~al\mbox{.}(2019)]%
        {Tillet2019}
\bibfield{author}{\bibinfo{person}{Philippe Tillet}, \bibinfo{person}{H.~T.
  Kung}, {and} \bibinfo{person}{David Cox}.} \bibinfo{year}{2019}\natexlab{}.
\newblock \showarticletitle{Triton: an intermediate language and compiler for
  tiled neural network computations}. In \bibinfo{booktitle}{\emph{Proceedings
  of the 3rd ACM SIGPLAN International Workshop on Machine Learning and
  Programming Languages}} \emph{(\bibinfo{series}{PLDI ’19})}.
  \bibinfo{publisher}{ACM}.
\newblock
\urldef\tempurl%
\url{https://doi.org/10.1145/3315508.3329973}
\showDOI{\tempurl}


\bibitem[Trott et~al\mbox{.}(2022)]%
        {Trott2022}
\bibfield{author}{\bibinfo{person}{Christian~R. Trott}, \bibinfo{person}{Damien
  Lebrun-Grandie}, \bibinfo{person}{Daniel Arndt}, \bibinfo{person}{Jan
  Ciesko}, \bibinfo{person}{Vinh Dang}, \bibinfo{person}{Nathan Ellingwood},
  \bibinfo{person}{Rahulkumar Gayatri}, \bibinfo{person}{Evan Harvey},
  \bibinfo{person}{Daisy~S. Hollman}, \bibinfo{person}{Dan Ibanez},
  \bibinfo{person}{Nevin Liber}, \bibinfo{person}{Jonathan Madsen},
  \bibinfo{person}{Jeff Miles}, \bibinfo{person}{David Poliakoff},
  \bibinfo{person}{Amy Powell}, \bibinfo{person}{Sivasankaran Rajamanickam},
  \bibinfo{person}{Mikael Simberg}, \bibinfo{person}{Dan Sunderland},
  \bibinfo{person}{Bruno Turcksin}, {and} \bibinfo{person}{Jeremiah Wilke}.}
  \bibinfo{year}{2022}\natexlab{}.
\newblock \showarticletitle{Kokkos 3: Programming Model Extensions for the
  Exascale Era}.
\newblock \bibinfo{journal}{\emph{{IEEE} Transactions on Parallel and
  Distributed Systems}} \bibinfo{volume}{33}, \bibinfo{number}{4}
  (\bibinfo{date}{apr} \bibinfo{year}{2022}), \bibinfo{pages}{805--817}.
\newblock
\urldef\tempurl%
\url{https://doi.org/10.1109/tpds.2021.3097283}
\showDOI{\tempurl}


\bibitem[{\DJ}uki{\'{c}} and Mi{\v{s}}i{\'{c}}(2023)]%
        {Dukic2023}
\bibfield{author}{\bibinfo{person}{Jovan {\DJ}uki{\'{c}}} {and}
  \bibinfo{person}{Marko Mi{\v{s}}i{\'{c}}}.} \bibinfo{year}{2023}\natexlab{}.
\newblock \showarticletitle{An Evaluation of Directive-Based Parallelization on
  the {GPU} Using a Parboil Benchmark}.
\newblock \bibinfo{journal}{\emph{Electronics}} \bibinfo{volume}{12},
  \bibinfo{number}{22} (\bibinfo{date}{nov} \bibinfo{year}{2023}),
  \bibinfo{pages}{4555}.
\newblock
\urldef\tempurl%
\url{https://doi.org/10.3390/electronics12224555}
\showDOI{\tempurl}


\bibitem[Usha et~al\mbox{.}(2020)]%
        {Usha2020}
\bibfield{author}{\bibinfo{person}{R. Usha}, \bibinfo{person}{Prachi Pandey},
  {and} \bibinfo{person}{N. Mangala}.} \bibinfo{year}{2020}\natexlab{}.
\newblock \showarticletitle{A Comprehensive Comparison and Analysis of
  {OpenACC} and {OpenMP} 4.5 for {NVIDIA} {GPUs}}. In
  \bibinfo{booktitle}{\emph{2020 {IEEE} High Performance Extreme Computing
  Conference ({HPEC})}}. \bibinfo{publisher}{{IEEE}}.
\newblock
\urldef\tempurl%
\url{https://doi.org/10.1109/hpec43674.2020.9286203}
\showDOI{\tempurl}


\bibitem[Ólason et~al\mbox{.}(2021)]%
        {Olason2021}
\bibfield{author}{\bibinfo{person}{Einar Ólason}, \bibinfo{person}{Pierre
  Rampal}, {and} \bibinfo{person}{Véronique Dansereau}.}
  \bibinfo{year}{2021}\natexlab{}.
\newblock \showarticletitle{On the statistical properties of sea-ice lead
  fraction and heat fluxes in the Arctic}.
\newblock \bibinfo{journal}{\emph{The Cryosphere}} \bibinfo{volume}{15},
  \bibinfo{number}{2} (\bibinfo{date}{Feb.} \bibinfo{year}{2021}),
  \bibinfo{pages}{1053--1064}.
\newblock
\showISSN{1994-0424}
\urldef\tempurl%
\url{https://doi.org/10.5194/tc-15-1053-2021}
\showDOI{\tempurl}


\end{thebibliography}

\appendix

\section{Code}\label{sec:appendixcode}
\begin{figure*}
\begin{lstlisting}[numbers=left, caption={Implementation of the stress update with Eigen. The method is generic in the degrees of freedom of the different cG and dG elements. The \lstinline|matrix()| and \lstinline|array()| methods change the type of an expression to differentiate between matrix and component-wise operations and are no-ops during runtime.}, ,captionpos=b, label=code:stresscpp]
template<DG> using DGVec = Eigen::Matrix<T, Eigen::Dynamic, DG>;

template <int CG, int DGstress, int DGadvection>
void StressUpdateHighOrder(const VPParameters& vpparameters, 
	const ParametricMomentumMap<CG>& pmap, const ParametricMesh& smesh,
	DGVec<DGstress>& S11, DGVec<DGstress>& S12, DGVec<DGstress>& S22, 
	const DGVec<DGstress>& E11, const DGVec<DGstress>& E12, const DGVec<DGstress>& E22,
	const DGVec<DGadvection>& H, const DGVec<DGadvection>& A, double alpha, double beta)
{
	constexpr int NGP = ((DGstress == 8) || (DGstress == 6)) ? 3 : (DGstress == 3 ? 2 : -1);
	using EdgeVec = Eigen::Matrix<T, 1, NGP * NGP>;

#pragma omp parallel for
	for (size_t i = 0; i < smesh.nelements; ++i) {
		auto hGauss = (H.row(i) * PSI<DGadvection, NGP>).array().max(0.0).matrix();
		auto aGauss = (A.row(i) * PSI<DGadvection, NGP>).array().max(0.0).min(1.0).matrix();
		EdgeVec P = (_vpparameters.Pstar * hGauss.array()
			* (-20.0 * (1.0 - aGauss.array())).exp()).matrix();
		
		const EdgeVec e11Gauss = E11.row(i) * PSI<DGstress, NGP>;
		const EdgeVec e12Gauss = E12.row(i) * PSI<DGstress, NGP>;
		const EdgeVec e22Gauss = E22.row(i) * PSI<DGstress, NGP>;
		const auto DELTA = (vpparameters.DeltaMin * vpparameters.DeltaMin
			+ 1.25 * (e11Gauss.array().square() + e22Gauss.array().square())
			+ 1.50 * e11Gauss.array() * e22Gauss.array() 
			+ e12Gauss.array().square()).sqrt().matrix();
		
		const T alphaInv = 1.0 / alpha;
		const T fac = 1.0 - alphaInv;
		const EdgeVec PDelta = P.array() / DELTA.array();	
		S11.row(i) = fac * S11.row(i) + (pmap.iMJwPSI[i] 
			* (alphaInv * (PDelta.array()
				* ((5.0 / 8.0) * e11Gauss.array() + (3.0 / 8.0) * e22Gauss.array())
				- 0.5 * P.array()).matrix().transpose())).transpose();
		S12.row(i) = fac * S12.row(i) + (pmap.iMJwPSI[i]
			* (alphaInv * (PDelta.array() * (1.0 / 4.0) * e12Gauss.array())
			.matrix().transpose())).transpose();
		S22.row(i) = fac * S22.row(i) + (pmap.iMJwPSI[i]
			* (alphaInv * (PDelta.array()
				* ((5.0 / 8.0) * e22Gauss.array() + (3.0 / 8.0) * e11Gauss.array())
				- 0.5 * P.array()).matrix().transpose())).transpose();
	}
}		
\end{lstlisting}
\end{figure*}


\end{document}